\documentclass[aps,prl,twocolumn,showpacs,superscriptaddress,amssymb,floatfix,tightenlines]{revtex4}

\usepackage{amsmath}
\usepackage{epsfig}
\usepackage{graphicx}
\usepackage{dcolumn}
\usepackage{bm}

\newcommand{\ket}[1]{\left| #1 \right\rangle}
\newcommand{\bra}[1]{\left\langle #1 \right|}

\newcommand{\av}[1]{\langle #1\rangle}

\newcommand{\raiz}{\mbox{$\textstyle \frac{1}{\sqrt{2}}$}}
\newcommand{\w}{\omega}

\begin{document}

\title{Radiative corrections and quantum gates in molecular systems}
\author{John H. Reina}
\affiliation{Centre for Quantum Computation,  Physics Department,
 Oxford University, OX1 3PU, United Kingdom\\
Materials Department, Oxford University, Oxford OX1 3PH, United
Kingdom}
\altaffiliation[On leave from ]{Centro
Internacional de F\'isica, A.~A.~4948, Bogot\'a, Colombia~(j.reina-estupinan@physics.ox.ac.uk).}
\author{Ray G. Beausoleil}
\affiliation{Hewlett-Packard Laboratories, 13837 175$^\textrm{th}$
Pl.\ NE, Redmond, WA 98052--2180, USA}
\author{Tim P. Spiller}
\affiliation{Hewlett-Packard Laboratories, Filton Road, Stoke
Gifford,
 Bristol BS34 8QZ, United Kingdom}
\author{William J. Munro}
\affiliation{Hewlett-Packard Laboratories, Filton Road, Stoke
Gifford, Bristol BS34 8QZ, United Kingdom}

\date{\today}

\begin{abstract}
We propose a method for quantum information processing using
molecules coupled to an external laser field. This utilizes
molecular interactions, control of the external field and an effective energy
shift of the doubly-excited state of two coupled molecules. Such a
level shift has been seen in the two-photon resonance experiments
recently reported in Ref.~\cite{hettich}. Here we show
that this can be explained in terms of the QED Lamb shift. 
We quantify the performance of 
the proposed quantum logic gates in the presence of dissipative mechanisms. 
The unitary transformations required for performing one- and two-qubit operations can be
implemented with present day molecular technology. The proposed techniques can
also be applied to coupled quantum dot and biomolecular systems. 
\end{abstract}

\pacs{03.67.Lx, 42.50.Fx, 32.80.Qk, 33.50.Dq}

\maketitle

There is currently enormous interest in the field of quantum
information processing (QIP). Potentially useful
applications range from speeding up computations such as
factoring, which would require tens of thousands of qubits,
through searching and quantum simulations, which could be useful
with a hundred or less, down to teleportation and quantum
repetition of communication, which are effectively just few-qubit
QIP. There have been some successful demonstrations of few qubit
manipulations, and there exists a vast and still-growing range of
proposals for realising QIP. The way forward is still very
open---the key routes to few-qubit and large-scale QIP, which
could well differ, have yet to be identified. Of particular
interest are new proposals that are based on already observed
phenomena, or that are clearly amenable to experimental
investigation with current technology. 

In this Letter we show how to
use small quantum networks of coupled molecules to implement gate
operations that enable universal QIP. A crucial
ingredient to this is an effective energy shift of the doubly-excited state
of two coupled molecules. 
We show that this can be explained in terms of the 
Lamb shifts of the levels, and we discuss
how such a system can be tailored, with current technology, to
realise a universal gate-set. We quantify the gate performance, and
calculate molecular resonance fluorescence spectra for comparison with
experiment. Although the physical origin of the relevant energy level shift
will be (potentially) different, it is clear that the 
approach can also be applied to other systems, such as coupled self-assembled
quantum dots or biomolecules. 

We commence by describing our set-up. Consider  two two-level
closely spaced molecules fixed at positions $\mathbf{r}_i$, which
have excitation frequencies  $\omega_{i}$, $i=1,2$, and are
separated by the vector $\mathbf{r}_{12}$. The 
dipole-coupled molecules are embedded in a dispersive
medium of refraction index $n$ and interact with the quantized
radiation field and with an external coherent driving field of
frequency $\omega_L$.  Evidence of such dipole-dipole ($d\textrm{-}d$) coupling
and the generation of sub and super-radiant states  has recently
been experimentally  reported in a set-up that involves single
terrylene  molecules  in a para-terphenyl crystal~\cite{hettich}.
We denote the ground and excited molecular states by
$\ket{g_i}\equiv \ket{0_i}$, and  $\ket{e_i}\equiv \ket{1_i}$,
with associated   transition dipole moments $\mathbf{d}_{i}$, and
corresponding spontaneous emission rates $\Gamma_i$. 
The strength
of the molecule-laser coupling is given by $\ell_i\equiv
-\mathbf{d}_{i}\cdot \mathbf{E}_i$, with $\mathbf{E}_i$ being the
amplitude of the coherent driving field at  $\mathbf{r}_i$. 
The coupled molecules can be represented as a single four-level 
system in \textsc{SU(4)};
we write the  system interaction Hamiltonian
in the computational basis of direct product states $\ket{i}\otimes\ket{j}$ $(i, j=0,1)$ as
\begin{equation}
\label{gralrpn}
\widehat{H}  = [\hat{H}_0^{(1)}+\hat{H}_L^{(1)}]\otimes\mathbb{\hat{I}}^{(2)}+
\mathbb{\hat{I}}^{(1)} \otimes[\hat{H}_0^{(2)}+\hat{H}_L^{(2)}]+\hat{H}_{12} \ ,
\end{equation}
where $hV_{12}$ denotes the $d\textrm{-}d$ interaction energy, and
the free, laser, and interaction Hamiltonians are given by
$\hat{H}_0^{(i)}=\frac{\hbar\omega_{i}}{2}[\hat{P}_{22}^{(i)}-\hat{P}_{11}^{(i)}]$, $\hat{H}_L^{(i)}=\hbar\ell^{(i)}[\hat{P}_{12}^{(i)}e^{i\w_Lt}+\hat{P}_{21}^{(i)}e^{-i\w_Lt}]$,
$\hat{H}_{12}=\hbar V_{12}[\hat{P}_{21}^{(1)}\otimes
\hat{P}_{12}^{(2)}]+\hbar V_{12}^\ast[\hat{P}_{12}^{(1)}\otimes
\hat{P}_{21}^{(2)}]$, with $\hat{P}_{ij}=\ket{i}\bra{j}$ denoting
the transition operators, and $\ket{i}$ and $\ket{j}$ being the
eigenvectors spanning the Hilbert space. One can then map the
transition operators into the pseudo-spin-1/2 dipole raising and
lowering operators $\hat{S}_i^\pm$:
$\hat{S}_i^+=\hat{P}_{21}^{(i)}$,
$\hat{S}_i^-=\hat{P}_{12}^{(i)}$,
$\hat{S}_i^+\hat{S}_i^-=\hat{P}_{22}^{(i)}$, and
$\hat{S}_i^-\hat{S}_i^+=\hat{P}_{11}^{(i)}$, thus generating the
corresponding operator algebra in \textsc{SU(2)$\otimes$SU(2)}. We
allow for a detuning  $\Delta_i=\omega_i-\omega_{L}$ between the
incident field and the molecular transitions.
We also
incorporate a shift $\Delta_\epsilon$ of the doubly-excited
state in our model---we shall show that this shift is an important
feature of a robust molecular quantum gate.
In our work we encode two qubits into this two-molecule system. Related work
on such systems,
where a single qubit is encoded into a molecular dimer, is
discussed in reference \cite{petrosyan02}.
In the
following, we make use of the electric-dipole approximation $k_0
r_{12}\ll 1$ ($ck_0\equiv 2\pi\nu_0=2\pi(\nu_1+\nu_2)/2$) and the
detuning parameters $\Delta_-=\omega_1-\omega_2$,
$\Delta_+=\omega_1+\omega_2-2\omega_L$~\cite{note1}.

\begin{figure}[!htb]
\includegraphics[width=8.7cm,height=6cm]{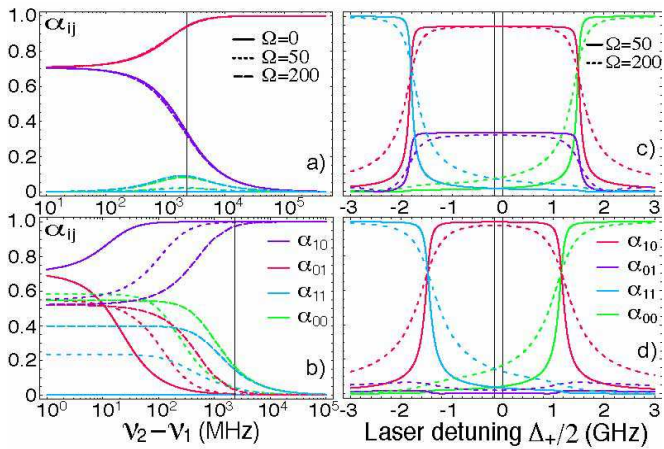}
  \caption{Eigenstate coefficients $\alpha_{ij}$ of the coupled 
system in the absence and presence of laser pumping
 as a function of $\Delta_-$, and  detuning $\Delta_+/2$. 
The weights of the basis states of the  4-level system have been 
plotted for $V_{12}=950 \, \textrm{MHz}$ (graphs (a) and (c)) 
and for  $V_{12}=10 \, \textrm{MHz}$  
((b) and (d)). $\Delta_-=2320 \, \textrm{MHz}$ in graphs (c) and (d). 
Notice the difference in eigenstates  behaviour  
for $\Delta_-\sim V_{12}$, and  $\Delta_-/V_{12}\gg 1$~\cite{note4}.}
\label{fidelity0}
\end{figure}

The eigenstate coefficients of Hamiltonian~(\ref{gralrpn}) are
depicted in Fig.~\ref{fidelity0} in terms of the different control
parameters of the model. We have used some of the  experimental
data reported in~\cite{hettich} and some other input values that
reveal interesting scenarios.
The eigenenergy
spectra (not shown) indicate that indeed the energy difference
between the entangled states shown in Fig.~\ref{fidelity0}(a), for
$\Delta_-=2320 \, \textsc{MHz}$ (vertical line), is about 3 GHz, in
agreement with the spectral results of~\cite{hettich}. 
However,
notice that as the laser coupling is increased,  there is a more
significant  contribution from all the basis states to the system
eigenstates, thus producing (e.g., for
$\Delta_-=2320 \, \textsc{MHz}$) the superpositions
$\alpha_{00}\ket{00}+\alpha_{01}\ket{01}+\alpha_{10}\ket{10}+\alpha_{11}\ket{11}$
with weight  coefficients $\alpha_{ij}$ as illustrated in
Fig.~\ref{fidelity0}(a). This means that for strong laser pumping,
$\alpha_{00}$ and $\alpha_{11}\neq 0$ in contrast to the entangled
sub- and super-radiant states reported in~\cite{hettich}. As shown
in Fig.~\ref{fidelity0}, such entangled states $\ket{\Psi_a}$ and
$\ket{\Psi_s}$~\cite{note4}  can indeed be created but for a
different parameter window (e.g., for $\Delta_-/V_{12}>500$). 
Figure~\ref{fidelity0} also shows that, for a fixed $V_{12}$, the
bigger the energy difference $\w_1-\w_2$, the easier one can achieve a `disentangling
effect', where the system's eigenstates  can become, to a very
good approximation, equal to  the computational basis states. This
is illustrated as a function of the detuning parameter in
Figs.~\ref{fidelity0}(c) and (d). 

As a result of the dipole-dipole interaction between molecules, 
the system exhibits resonant cooperative absorption of light~\cite{agarwal}, 
thus  allowing simultaneous excitation by absorption of two-photons; 
we represent the spontaneous  decay of such excitations by the rate  
$\Gamma_{12}=\Gamma_{21}$. It is not difficult to calculate the general 
expressions for the $d\textrm{-}d$ interaction strength and the incoherent 
decay rate $\Gamma_{12}$~\cite{note3}. In the near field ($r_{12}\ll\lambda_L$) 
the interaction energy and the incoherent decay rate are given by  
$V_{12}= \frac{3\sqrt{\Gamma_1 \Gamma_2}}{8\pi z^3}[\hat{\mathbf{d}}_{1}\cdot  \hat{\mathbf{d}}_{2}
-3(\hat{\mathbf{d}}_{1}\cdot
      \hat{\mathbf{r}}_{12})
(\hat{\mathbf{d}}_{2}\cdot
      \hat{\mathbf{r}}_{12}) ] $, and $\Gamma_{12}=\sqrt{\Gamma_1 \Gamma_2}\, 
\hat{\mathbf{d}}_{1}\cdot  \hat{\mathbf{d}}_{2}
$, and hence the maximum (minimum) strengths are obtained for
parallel (perpendicular) dipole moments.  $z=nk_{0}r_{12}$,
$\Gamma_i=\frac{nA_{12;i}}{2}=n\omega_i^3\|\mathbf{d}\|^2/(3\epsilon_0
hc^3)$,  where $A_{12;i}$ ($i=1,2$) are the Einstein coefficients
in vacuum, and $ \hat{\mathbf{d}}_{i}$ and $\hat{\mathbf{r}}_{12}$
are unit vectors along 
$\mathbf{d}_{i}$ and along
$\mathbf{r}_{12}$.  
The presence of the
dielectric modifies the molecules radiative decay rates which in
turn implies a shift in the molecular energy levels.  
The vacuum frequencies are now modified and hence the phase
velocity of light waves in the dielectric medium is given by
$c/n(\omega)=\omega/k(\omega)$.

\begin{figure}[!htb]
\begin{center}
\includegraphics[width=7.0cm,height=7.0cm]{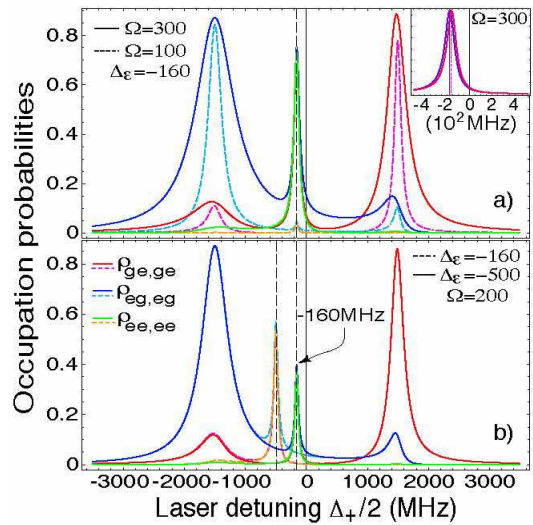}
\caption{Steady-state density matrix elements as a function of the
laser detuning parameter $\Delta_+/2$.  The  occupation
probabilities give the structure of the resonance fluorescence
spectrum, as is shown in a) for different excitation intensities
$\ell_i$, and in b) for different effective shifts $\Delta_\epsilon$;
$\ell_i\equiv \Omega$, $\Gamma_i=50 \, \textrm{MHz}$,
$\Gamma_{12}=9 \, \textrm{MHz}$, $\Delta_-/2=1160 \,
\textrm{MHz}$, and $V_{12}=950 \, \textrm{MHz}$.} 
\label{fluor}
\end{center}
\end{figure}

We describe the full molecular quantum dynamics by solving the
master equation for the  Lindblad operator~\cite{note5}, thus
taking into account all the possible excitation decay channels
$\Gamma_i$, and $\Gamma_{ij}$. 
In particular, Ref.~\cite{hettich}  reports  an energy shift  in
the fluorescence spectrum associated with the upper level
$E_\epsilon$ of $\Delta_+/2\simeq -160 \, \textrm{MHz}$ 
(see the
central resonance of Fig.~\ref{fluor}(a)), where the  steady-state
density matrix element $\rho_{ee,ee}\equiv \av{e_1e_2|  \rho |e_1e_2}$ displays the observed shift
($\Delta_+/2=-160 \, \textrm{MHz}$, vertical dashed line). 
This effective shift can be attributed to the Lamb shifts of
the three states  $\ket{\Psi_{a}}$, $\ket{\Psi_{s}}$ and $\ket{\Psi_{\epsilon}}$.
Decay channel interference modifies the 
decay rates of the first two states \cite{hettich}
(and hence their Lamb shifts) away from the uncoupled values. As a consequence,
the doubly-excited state is effectively displaced from where it would be expected,
given the energies of the lower two levels.
The Lamb shifts can be calculated by taking into account
self-interactions and by renormalizing to remove quadratic
divergences~\cite{agarwal,milonni} from
$\Lambda_{i}=-\frac{\|\mathbf{d}_{i}\|^2\omega_i^3}{3\pi
\epsilon_0 hc^3}\int
n(\omega)\left[(\omega+\omega_i)^{-1}+(\omega-\omega_i)^{-1}\right]d\omega$. 
A detailed calculation for  the experimental set-up reported in~\cite{hettich} 
shows that  
$\Delta_{\epsilon} = \Lambda_{\epsilon}-\Lambda_{s}-\Lambda_{a} \simeq  -178 \, 
\textrm{MHz}$, 
a result that is in good agreement with the observed value, as is shown by
the inset of Fig.~\ref{fluor} where the fluorescence around this 
excitation energy is plotted for  both the theoretical (solid line)
and the measured (dashed line) values as a function of the laser detuning $\Delta_+/2$.
As we show below, the $\Delta_{\epsilon}$ shift is clearly useful for QIP; 
even more so because it can be
tailored by means of  an appropriate engineering of 
the structural
and optical properties of the  molecular set-up. 
Different strengths $\Delta_\epsilon$
are shown in
Fig.~\ref{fluor}(b)
where the expected fluorescence spectra 
is plotted as a function of the laser
detuning $\Delta_+/2$.
The result that additional
energy shifts appear in the 
spectra is viewed here as a
meaningful resource for performing molecular two-qubit logic gates
and therefore for quantum computing purposes. Although we have
focused on the experimental data reported in~\cite{hettich}, other
parameter windows and different molecular systems can certainly be
explored in the light of our results.

The splittings in the energy spectrum of the interacting system can be used  to  induce a conditional dynamics between the two subsystems (e.g.,  a laser with $\lambda_L\simeq 578\,\textrm{nm}$ and linewidth $\Delta \nu\simeq 1\,\textsc{MHz}$ can be suitable~\cite{hettich}).  Figure~\ref{fidelity0}   displays the eigenstate coefficients  in the absence and presence of laser pumping for different interaction strengths $V_{12}$.  For  $\ell_i=0$,  the energy separation between $\ket{\Psi_s}$ and
$\ket{\Psi_a}$   is  $\sqrt{\Delta_-^2+4V_{12}^2}$~\cite{note4} and hence
molecules of equal transition energies ($\Delta_-=0$) have $c_1=c_2=1$ and the corresponding symmetric and antisymmetric states  become the maximally entangled states (MES) $\raiz(\ket{e_1g_2} \pm\ket{g_1e_2})$, as can also be seen in Figs.~\ref{fidelity0}(a) and (b). Otherwise ($\omega_1\neq \omega_2$), the degree of molecular entanglement can be manipulated by controlling the ratio
$V_{12}/\Delta_-$. This can be experimentally  done by means of applying a differential Stark shift via an external inhomogeneous  electric field to select different values of the detuning  $\Delta_-$  or by directly tuning the dipole-dipole interaction by means of changing the separation $r_{12}$ and/or by using a medium with different dispersive  properties. This means that a rich spectra of entangled symmetric and antisymmetric states can be generated by modifying the eigenstate coefficients. It follows from~\cite{note4} that
$\alpha_1 = \sqrt{(Y+1)/2Y}$, $\alpha_2 = \sqrt{(Y-1)/2Y}$ ($Y\equiv X/\Delta_-$),  then,  if the system's ratio $V_{12}/\Delta_-\ll 1$, $\alpha_2\approx V_{12}/\Delta_-$ ($c_i/\sqrt{2}\equiv \alpha_i$):
the bigger the difference  in the transition energies (compared to $V_{12}$)
the more disentangled the intermediate states become.  

\begin{figure}[!htb]
\begin{center}
\includegraphics[width=7.5cm]{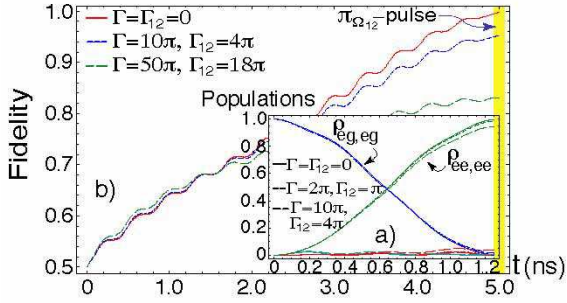}
\caption{(a) $\textsc{cnot}$ gate operation: a $\pi_{\Omega_{12}}$-pulse is applied  
in order to generate the conditional logic $\ket{e_1g_2}\mapsto \ket{e_1e_2}$. 
The corresponding populations are shown as a function of time in the absence and presence of 
dissipative channels.
$\Delta_+/2=-1319\, \textrm{MHz}$, $\Delta_-=-2320\, \textrm{MHz}$, $\ell_i\equiv \Omega=200\, \textrm{MHz}$, $\Delta_\epsilon=160 \, \textrm{MHz}$, and $V_{12}=50 \, \textrm{MHz}$. (b) Fidelity (see text) in the generation of the MES
$\raiz (\ket{g_1g_2}+\ket{e_1e_2})$. The bar signals the end of the applied $\pi_{\Omega_{12}}$-pulse and indicates a measure of the actual entangled state achieved  in the absence and presence of dissipation. Input parameters are as before except for  $\Omega=50 \, \textrm{MHz}$.
Decay rates are given in $10^6 \, \textrm{s}^{-1}$ units.} 
\label{fidelity}
\end{center}
\end{figure}

The effective shift $\Delta_\epsilon$ described above
implies that the resonant frequency for transitions between the
basis states $\ket{g_i}$ and $\ket{e_i}$ of one qubit \emph{depends}
on the state of the neighbouring qubit: the energy difference
between states $\ket{\Psi_a}$ and that of $\ket{\Psi_g}$ is
different from the one between states $\ket{\Psi_\epsilon}$ and
$\ket{\Psi_s}$ due to $\Delta_\epsilon$ and $V_{12}$~\cite{note4}.
Thus, in the scenario $V_{12}/\Delta_-\ll 1$, we can naturally construct the two-qubit
\textsc{cnot} gate
$U_\textsc{cnot}$: $\ket{m}\otimes\ket{n}\mapsto\ket{m}\otimes\ket{m\oplus n}$.
Hence, the logic operation
$\ket{e_1}\ket{g_2}\mapsto \ket{e_1}\ket{e_2}$ can be achieved by illuminating the input qubit system with a $\pi$-pulse of energy  $\Omega_{12} = \w_2-\delta+\Delta_\epsilon$ (say, a
$\pi_{\Omega_{12}}$-pulse), where $\delta=V_{12}^2/\Delta_-$.
This is illustrated, for experimentally attainable values~\cite{hettich}, in Fig.~\ref{fidelity}(a) where the system's initial state $\ket{e_1g_2}$ is induced to evolve onto the final state $\ket{e_1e_2}$ via the application of a $\pi_{\Omega_{12}}$-pulse, thus realizing a quantum $\textsc{cnot}$ gate in 1.25 ns. In the absence of dissipative channels the gate operation exhibits a perfect fidelity (see below), whereas the inclusion of realistic decay rates shows the appearance of small weight contributions arising from the populations $\rho_{gg,gg}$ and $\rho_{ge,ge}$ (curves at the bottom of Fig.~\ref{fidelity}(a)). This, however, can be further controlled by means of choosing different  $\Delta_\epsilon$, $V_{12}/\Delta_-$, and laser strengths to improve the gate fidelity.
Conversely, if the role of the control qubit is to be performed by the second qubit, the gate operation
$\ket{g_1}\ket{e_2}\mapsto \ket{e_1}\ket{e_2}$ can be realized via a
$\pi_{\Omega_{21}}$-pulse  of frequency $\Omega_{21} = \w_1+\delta+\Delta_\epsilon$.
The energy selectivity of such a \textsc{cnot} gate is thus determined by
$\Delta_-$, $\Delta_\epsilon$ and $V_{12}$, and typical operation time scales can be gathered from Figs.~\ref{fluor} and \ref{fidelity}  for different  parameter windows. The complimentary  unitaries  in \textsc{SU(2)} required to build a  universal set of gates can be achieved  by inducing appropriate Rabi oscillations of the corresponding dipole operator expectation value $\av{\mathbf{\hat{d}}}=\av{\hat{P}_{12}\mathbf{d}_{12}+\hat{P}_{21}\mathbf{d}_{21}}$. Thus, single qubit control can be performed by tailoring the energy and length of applied laser pulses to induce distinct Rabi oscillations.

The scenario considered previously exploits the diagonal shifts $\Delta_\epsilon$ by means of focusing on a particular  parameter window. However, even for arbitrary $V_{12}/\Delta_-$ ratios one can still perform entangling gates that enable universal QC. By writing the interaction Hamiltonian of Eq.~(\ref{gralrpn}) as $\hat{H}_{12}=\hbar V_{12}[\hat{\sigma}_x\otimes\hat{\sigma}_x+\hat{\sigma}_y\otimes\hat{\sigma}_y]\equiv\hat{H}_{xy}$, one can define an \emph{optimal } entangling  gate $U_{xy}$ for the $d\textrm{-}d$ interaction by simply letting the system evolve under  $\hat{H}_{xy}$ for a time $t_{xy}=\pi/4V_{12}$. By means of the  self-adjoint operator $\hat{H}_{\vec{\mu}}=\sum_i\mu_i\hat{\sigma}_i\otimes\hat{\sigma}_i$, $\hat{U}_{\vec{\mu}}\equiv \exp(-i\hat{H}_{\vec{\mu}})$ provides, up to \textsc{LU}, a decomposition for \emph{any } two-qubit gate and thus  the $d\textrm{-}d$ interaction 
has associated 
$U_{xy}$: $\ket{m}\otimes\ket{n}\mapsto i^{|m-n|}\ket{n}\otimes\ket{m}$ as its `natural' entangling gate~\cite{hammerer02}. This gate,  equivalent to a combination of  \textsc{cnot} and  \textsc{swap}, plus \textsc{LU}, are a universal gate-set.  This means that, provided one has access to performing sufficiently fast \textsc{LU} operations, one could always  exploit the $d\textrm{-}d$ interaction for universal QC.
Typical switching times for the experimental data analyzed here can be in the sub-picosecond time scale ($\sim$1/1000 the fastest decay rate).

We calculate the steady-state density matrix elements in the computational basis as a function of the laser detuning $\Delta_+/2$.
Fig.~\ref{fluor}(a) shows a strong enhancement  of the  two-photon resonance (TPR) as the strength of the laser coupling is increased (for  $V_{12}=950 \, \textrm{MHz}$). As the $d\textrm{-}d$ coupling determines the size of the TPR, the weaker the interaction the more difficult the detection of the TPR fluorescence peak.
Fig.~\ref{fluor}(b) shows effectively that  an increase in $\Delta_\epsilon$ has  associated with it a further TPR shift that could then be observed in the fluorescence spectrum, as predicted above. The asymmetric shape of the resonance lines arises from the destructive and constructive interference between the system's excitation decay channels $\ket{\Psi_g}\mapsto \ket{\Psi_a}\mapsto\ket{\Psi_\epsilon}$, and $\ket{\Psi_g}\mapsto \ket{\Psi_s}\mapsto\ket{\Psi_\epsilon}$, which in turn is also reflected in the observed linewidth difference between the left and right peaks (Fig.~\ref{fluor}), a hallmark of Dicke's subradiance  and superradiance.

The quality of the gates discussed here depends on the
ability to control the relevant structural and external  parameters.
An indication of gate performance is the quality of the maximally entangled states they produce.
Due to decoherence, any attempt to generate a MES
$\ket{\psi_{m}}$ in practice creates a mixed state $\rho$.  
In Fig.~\ref{fidelity}(b) we plot the corresponding
fidelity $\mathcal{F}\equiv \sqrt{\av{\psi_{m}|  \rho | \psi_{m}}}$ for $\ket{\psi_{m}} \equiv \raiz (\ket{g_1g_2}+\ket{e_1e_2})$. 
The initial state is taken to be $\raiz (\ket{g_1}+\ket{e_1})\ket{g_2}$ (this can be achieved by starting in the system's ground state and then applying a $\pi/2$  pulse at energy $\w_1$), and hence $\ket{\psi_{m}}$ can be generated by applying a $\pi_{\Omega_{12}}$-pulse.  
Figure~\ref{fidelity}(b) gives the fidelity between $\ket{\psi_{m}}$ and $\rho$ in the presence of dissipative interactions. Clearly, zero decay gives a perfect fidelity ($\mathcal{F}=1$), but the action of dissipative processes (as described in~\cite{note5}) gives $\mathcal{F}\lesssim1$, as shown in Fig.~\ref{fidelity}(b) by the bar signaling the end of the applied $\pi_{\Omega_{12}}$-pulse. We have considered the rather non-optimistic ratio $\frac{V_{12}}{\Delta_-}\sim 2\times 10^{-2}$ but smaller ratios could indeed be experimentally achievable  and hence improved fidelities can be obtained in conjunction with tailored  $\Delta_\epsilon$, and laser field values.
Our numerical experiments consider  a setup and energy scales  that have been measured in the laboratory~\cite{hettich} thus providing a significant step  towards proof-of-principle experiments that can be undertaken with currently available technology.

Although of a different physical origin, we stress that the main
ideas discussed here can also be applied in the context of the
technologically important self-assembled coupled quantum dot
molecules~\cite{reina,crooker} and biomolecular systems~\cite{herek02}.
Summarizing, we have  given a detailed prescription for  molecular
conditional quantum dynamics that can be implemented with modest
experimental requirements  by tailoring control  parameters such
as excitation frequencies,  spontaneous decay rates,  interaction
strengths, and external laser pumping. Furthermore, we have explained 
the quantum mechanical origin of the TPR energy-level shift and 
we have highlighted its relevance to QIP protocols. We have shown that
the performance of a molecular quantum gate can be adjusted by
varying the value of this shift.

We thank V.~Sandoghdar for valuable correspondence.
JHR is supported by EPSRC as part of the Foresight LINK Award \emph{Nanoelectronics at the Quantum
Edge}.

\end{document}